\documentstyle[12pt,aasms4]{article}

\begin{document}

\title{The energetics of crystallizing white dwarfs revisited again}

\author{J. Isern}
\affil{Institut d'Estudis Espacials de Catalunya/CSIC, Edifici Nexus, 
       Gran Capit\`a 2--4, 08034 Barcelona, Spain}
\author{E. Garc\'{\i}a-Berro}
\affil{Departament de F\'{\i}sica Aplicada, Universitat Polit\`ecnica 
       de Catalunya, Jordi Girona Salgado s/n, M\`odul B--5, Campus 
       Nord, 08034 Barcelona, Spain}
\affil{Institut d'Estudis Espacials de Catalunya/UPC, Edifici Nexus, 
       Gran Capit\`a 2--4, 08034 Barcelona, Spain}
\author{M. Hernanz} 
\affil{Institut d'Estudis Espacials de Catalunya/CSIC, Edifici Nexus, 
       Gran Capit\`a 2--4, 08034 Barcelona, Spain}
\author{G. Chabrier}
\affil{Centre de Recherche Astrophysique de Lyon (UMR~CNRS~5574),
       Ecole Normale Superi\'erieure de Lyon, 69364 Lyon Cedex 07,
       France}

\received{}
\accepted{}

\begin{abstract}

The evolution of white dwarfs is a cooling process that depends 
on the energy stored in the core and on the way in which it is 
transferred through the envelope. In this paper we show that 
despite some (erroneous) claims, the redistribution of chemical 
elements ensuing the crystallization of C/O white dwarfs provides 
between the 10\% and the 20\% of the total energy released during 
the crystallization process, depending on the internal chemical 
composition, which is not negligible at all, given the present 
state of the art of the white dwarf cooling theory.

\end{abstract}


\section{Introduction}

The evolution of white dwarfs is essentially a cooling process. From 
this point of view, the star can be described as an energy reservoir 
(the inner core) surrounded by an opaque envelope (the outer partially 
degenerate layers) that controls the rate of cooling (Mestel, 1952). It 
is clear that an accurate description of the process demands a good 
knowledge of all the sources of energy as well as of the ``transparency 
degree'' of the atmosphere. 

One of the possible sources of energy is related with solidification. 
Crystallization not only induces a release of latent heat (Van Horn,
1968; Lamb \& Van Horn, 1975) but also a release of gravitational 
energy associated with a change of chemical composition during the 
phase transition (Stevenson, 1980; Mochkovitch, 1983; Isern et al.,
1991; Xu \& Van Horn, 1992). Although the importance of solidification 
of alloys was recognized long ago in geophysical circles, this is not 
yet the case in astrophysics and some misleading concepts and formulae 
are still appearing in the literature. For instance, it has been 
recently claimed (Hansen, 1999) that the amount of energy due to 
chemical differentiation at crystallization in white dwarf interiors 
is negligible, and much smaller than the latent heat contribution. 
This claim, that is incorrect, stems from an erroneous estimation 
of this differentiation contribution.

There are several factors that influence the computed delay introduced 
by the change of the chemical composition, namely: i) the phase diagram 
of the C/O binary mixture, ii) the chemical profile at the beginning 
of the white dwarf phase, and iii) the transparency of the envelope. 
Obviously, any change in any one of these inputs directly translates 
into a change of the computed delay, but the underlying physical 
picture is the same and the role played by this phenomenon has to 
be clarified if white dwarfs have to be used as effective tools for 
deriving important information about the history of the Galaxy.

In this paper we show how to correctly compute the release of 
gravitational energy associated to solidification (within the 
current limits of our knowledge of the physics of Coulomb plasmas) 
and we discuss the importance of this process for different 
assumptions about white dwarfs.

\section{The amount of energy released}

The local energy budget of the white dwarf can be written as (Isern 
et al., 1997):
\begin{equation}
-(\frac{dL_{\rm r}}{dm}+\epsilon_{\nu})=C_{\rm v}\frac{dT}{dt}+
T\Big(\frac{\partial P}{\partial T}\Big)_{V,X_0}\frac{dV}{dt}
-l_{\rm s}\frac{{dM}_{\rm s}}{dt}\delta(m-M_{\rm s})+
\Big(\frac{\partial E}{\partial X_0}\Big)_{T,V}\frac{dX_0}{dt}
\end{equation}
where $E$ is the internal energy per unit mass, $V=1/\rho$,
$l_{\rm s}$ is the latent heat of crystallization and $\dot{M}_
{\rm s}$ is the rate at which the solid core grows; the delta function
indicates that the latent heat is released at the solidification front. 
We are assuming that the white dwarf is made of two chemical species 
with atomic numbers $Z_0$ and $Z_1$, mass numbers $A_0$ and $A_1$ 
and abundances by mass $X_0$ and $X_1$, respectively ($X_0+X_1=1$, 
and the suffix 0 refers to the heaviest element). The chemical 
differentiation contributes to the luminosity not only through 
compressional work, which is almost negligible during the 
crystallization phase, but also through energy due to the 
change in the chemical abundances or, from a first-principle 
point of view, to a variation of the chemical potentials, which 
leads to the last term of this equation (Chabrier, 1997).
It is worthwhile to notice here that when the star has exhausted 
all the internal sources of energy, the compression of the outer 
layers prevents the star to disappear to undetectable luminosities 
(D'Antona \& Mazzitelli, 1990; Isern et al., 1997, 1998) and that a 
rigourous treatement of these layers is crucial to understand the 
properties of very cool white dwarfs.

A useful way to include the energetics of crystallization in 
numerical codes can be derived, under the assumption that the 
cooling process is slow enough, which is almost always the case. 
Then, a rate of release of energy per gram of crystallized matter, 
$\epsilon_{\rm g}$, related to the change in chemical composition, 
can be defined as follows (see equation (12) in Isern et al., 1997):
\begin{equation}
\epsilon_{\rm g}=-\Delta X_0
\Bigg[\Big(\frac{\partial E}{\partial X_0}\Big)_{M_{\rm s}}-
\Big\langle\frac{\partial E}{\partial X_0}\Big\rangle\Bigg] 
\end{equation}
where $\Delta X_0=X_0^{\rm sol}-X_0^{\rm liq}$. The first term 
represents the energy released in the layer that is crystallizing, 
as a consequence of the increase of concentration of the heaviest 
species, whereas the second term, in angular brackets, represents 
the energy that is absorbed in average in the convective layer that
appears just above the crystallization front, as a consequence of 
the decrease of concentration of the heaviest species.

In order to compute $(\partial E/\partial X_0)_{V,T}$, we should take 
into account that the internal energy per unit mass can be divided 
into the electronic and the ionic components. Although in Salaris 
et al. (1997) we used the complete expressions, here, for the sake of 
simplicity, we are only going to use the completely degenerate 
nonrelativistic expression for the electrons 
\begin{equation}
E_{\rm e}=\frac{3}{2}K_1\rho^{2/3}Y_{\rm e}^{5/3}
\end{equation} 
where $K_1=1.004\times 10^{13}$ (cgs units), and the ideal contribution
plus the Madelung term of the Coulomb energy for the ions
\begin{equation}
E_{\rm i}\simeq E_{\rm id}+E_{\rm Mad}\simeq 
\frac{\Re T}{\mu}\Big( \frac{3}{2} -0.9 \Gamma\Big)
\end{equation} 
where $\Re$ is the gas constant and $\Gamma=\Gamma_{\rm e}
\overline{Z^{5/3}}$ is the Coulomb coupling constant, with $\Gamma_
{\rm e}=2.272\times 10^5 (\rho Y_{\rm e})^{1/3}/T$ and 
$\overline{Z^{5/3}}=\mu [X_0 Z_0^{5/3}/A_0+(1-X_0) Z_1^{5/3}/A_1]$,
being $\mu$ the mean molecular weight per ion. Therefore, the 
derivative $(\partial E/\partial X_0)_{T,V}$ is the sum of three 
contributions:

\begin{equation}
\Big(\frac{\partial E_{\rm e}}{\partial X_0}\Big)_{T,V}=
\frac{5}{2} K_1\rho^{2/3}Y_{\rm e}^{2/3}
\Big( \frac{Z_0}{A_0} - \frac{Z_1}{A_1} \Big)
\end{equation}
\begin{equation}
\Big(\frac{\partial E_{\rm id}}{\partial X_0}\Big)_{T,V}=
\frac{3}{2} \Re T \Big( \frac{1}{A_0} - \frac{1}{A_1} \Big)
\end{equation}
\begin{eqnarray}
\Big(\frac{\partial E_{\rm Mad}}{\partial X_0}\Big)_{T,V}=
-0.9 \Re T \Gamma_{\rm e}&\Bigg[&
\frac{1}{3 Y_{\rm e}}
\Big( \frac{Z_0}{A_0} - \frac{Z_1}{A_1} \Big)
\Big( X_0 \frac{Z_0^{5/3}}{A_0}+(1-X_0)\frac{Z_1^{5/3}}{A_1} \Big)
\nonumber \\
&+& \Big( \frac{Z_0^{5/3}}{A_0}- \frac{Z_1^{5/3}}{A_1} \Big) \Bigg]
\end{eqnarray}

\noindent
Since the contributions to the electron mole number, $Y_{\rm e}$, of 
carbon and oxygen are very similar, the derivative $(\partial E/
\partial X_0)_{T,V}$ is dominated by the ionic contribution (and 
in particular by the second term of the Madelung term). It is 
important to remind that the electronic term (5) is not negligible 
when neutron-rich species are considered (i.e. $^{22}$Ne, $^{56}$Fe). 
In the case of C/O mixtures, we can retain only the second term in (7):

\begin{equation}
\Big(\frac{\partial E}{\partial X_0}\Big)_{T,V}\simeq
-0.9 \Re T \Gamma_{\rm e}
\Big(\frac{Z_0^{5/3}}{A_0}- \frac{Z_1^{5/3}}{A_1} \Big)
\end{equation}

\noindent
This expression can be directly substituted in equation (1), which
is well suited for numerical codes. Note as well that this term
when substituted in equation (1) is local and, therefore, generally 
speaking, can be considered as a source or a sink of energy, depending 
on the local change of chemical composition. 

In order to guess the importance of the process of redistribution 
related to crystallization, we can compare with the latent heat 
$l_{\rm s}\sim k_{\rm B}T$ per ion (Hansen, 1999). In this case, 
it is convenient to deal with the energy released per ion $\Delta 
u$:
\begin{equation}
\frac{\Delta u}{k_{\rm B}T} \simeq 0.9 \mu \Gamma_{\rm e}
\Big(\frac{Z_0^{5/3}}{A_0}-\frac{Z_1^{5/3}}{A_1} \Big)
\Delta X_0 \simeq \frac{\Delta u}{l_{\rm s}} 
\end{equation}

\noindent
Note that equations (8) and (9) are exactly the same as equations 
(7) and (8) of Chabrier (1997), where the energy is given explicitly 
per unit {\sl mass}, which implies the factor $\mu$ in the ratio
$\Delta u/k_{\rm B}T$. Typically, $\Delta X_0=0.2$, leading to 
$\Delta u/l_{\rm s}=6$, which is different from the 0.3 value quoted 
by Hansen (1999). The reason of this large difference is that in 
equation (1) of Hansen (1999), a factor $\mu$ is missing. Therefore, 
the energy released by the separation of carbon and oxygen is 
underestimated in Hansen (1999) by more than an order of magnitude, 
leading to erroneous conclusions.

Regarding the phase diagram of a C/O mixture, our first estimates were
based on the phase diagram of Stevenson (1980) that predicted complete
inmiscibility of carbon and oxygen in the solid phase. Using equation 
(2) and this phase diagram, the energy released by the chemical 
differentiation was $\Delta E_{\rm g}=6.6\times 10^{46}$ erg in 
the case of a 0.606 $M_{\sun}$ white dwarf with a flat initial 
chemical profile ($X_{\rm C}=X_{\rm O}=0.5$). This value has to 
be compared with the total energy released during the crystallization
phase --- which is equal to the difference $\Delta B$ of the binding 
energies --- when chemical differentiation is neglected, $\Delta B 
=7.2\times 10^{46}$ erg. Since then, the advances in the physics 
of dense plasmas led to a better understanding of the phase transition 
and the corresponding diagram turned out to be of the spindle form. 
Consequently, the degree of chemical separation was stronly reduced 
(Segretain et al., 1994). For the white dwarf with the aforementioned 
characteristics $\Delta E_{\rm g}=2.3\times 10^{46}$ erg is obtained, 
while the latent heat is $\Delta E_{\rm l}=1.8\times 10^{46}$ erg, 
which means that this effect is still significant. Since the latent 
heat has been computed assuming $l_{\rm s}\sim k_{\rm B}T$ per ion, 
the quoted value is, nevertheless, a gross estimate.

The initial chemical profile plays also a very important role (Hernanz 
et al., 1994; Segretain et al., 1994; Salaris et al., 1997). For 
instance, if a high effective rate of the $^{12}{\rm C}(\alpha,
\gamma)^{16}{\rm O}$ reaction is adopted, the abundance of oxygen 
in the central layers is as high as 0.74 by mass in the case of a 
0.606 $M_{\sun}$ and the degree of mixing in the liquid layers is 
strongly reduced by the very steep gradients of chemical composition 
that appear. In this case, the total gravitational energy release 
upon crystallization obtained using equation (2) is $\Delta E_{\rm g}
=1.1\times 10^{46}$ erg, which is still not negligible since it is of 
the same order as the latent heat. It is important to stress here that 
the complete version of equation (2) has to be used since the term in 
angular brackets, which depends on the degree of mixing of the liquid 
region, reduces the net gravitational energy released by the process. 
Moreover, the total energy released during the crystallization phase 
for the same white dwarf without introducing any simplification in 
the equation of state is $\Delta B=9.6\times 10^{46}$ erg when phase 
separation is included and $\Delta B=8.6\times 10^{46}$ erg when phase 
separation is neglected. The difference is caused by the contribution 
of the chemical redistribution process and confirms the values obtained 
above. Notice also that the chemical profiles adopted here are 
those that predict the maximum abundances of oxygen in the center 
and, consequently, minimize the effects of phase separation.

\section{The time delay in the cooling process}

According to equation (38) in Isern et al. (1997) or (11) in Chabrier 
(1997) the time delay in the cooling sequences can be expressed as 
\begin{equation}
\Delta t = \int ^{M_{\rm WD}}_0 \frac{\epsilon_{\rm g}(T_{\rm c})}
{L(T_{\rm c})} dm
\end{equation}
where $T_{\rm c}(m)$ is the core temperature when the crystallization
front is located at mass $m$ and $L$ is its corresponding luminosity. 
From this equation it is obvious that any change in the transparency 
of the envelope of the white dwarf and, thus, in the relationship 
between the luminosity and the core temperature, directly translates 
into the delay introduced by phase separation in the cooling times.
In order to see the influence of the envelope in the computed time 
delays we have adopted the following white dwarf model atmospheres. 
Our first model atmosphere has been obtained from the DA model sequence 
of Wood (1995), which has a mass fraction of the helium layer of 
$q_{\rm He}=10^{-4}$ and a hydrogen layer of $q_{\rm H}=10^{-2}$; 
the second model atmosphere is the non-DA model sequence of Wood 
\& Winget (1989) which has a helium layer of $q_{\rm He}=10^{-4}$. 
However it should be noted that between these two model sequences 
there was a substantial change in the opacities, and therefore the 
comparison is meaningless (i.e., the non-DA model is more opaque 
than the DA one). The remaining two model atmospheres are those of 
Hansen (1999) for both DA and non-DA white dwarfs. These atmospheres 
have been computed with state of the art physical inputs for both the 
equation of state and the opacities for the range of densities and 
temperatures relevant for white dwarf envelopes (although it should 
be mentioned that the contributions to the opacity of H$^{3+}$ and 
H$^{2+}$ ions were neglected in this calculation) and have the same 
hydrogen and helium layer mass fractions as those of Wood (1995) and 
Wood \& Winget (1989), respectively. For the sake of conciseness we 
will refer to the cooling sequences obtained with these two sets of 
envelopes as ``MW'' and ``BH'', respectively. In all the cases, the 
mass of the white dwarf is 0.606 $M_{\sun}$ and the initial chemical
profile of the C/O mixture is that of Salaris et al. (1997).

\placefigure{fig1}

It is interesting to compare the core temperature-luminosity 
relationship for both sets of sequences. This is done in figure 
1 where it can be seen that the DA model atmospheres MW and BH are 
in very good agreement down to temperatures of the order of $\log(
T_{\rm c})\simeq 6.5$, whereas at lower temperatures the model 
atmospheres BH predict significantly lower luminosities (that
is, they are less transparent). This is undoubtedly due to the 
significant improvement in cool white dwarf atmosphere calculations 
done in Hansen (1999). In contrast, the non-DA model atmosphere BH is 
by far more transparent at any temperature than the corresponding 
MW model. This is clearly due to the fact that the Wood \& Winget 
(1989) $L-T_{\rm c}$ relation was based on the old Los Alamos 
opacities which include a finite contribution from metals whereas 
Hansen's non-DA are pure helium. In any event, it is straightforward 
from our equation (10) to realize why the computed time delays are 
very different {\sl in absolute value} for the model cooling sequences 
reported here, as was clearly stated in Isern et al. (1997).

\placefigure{fig2}

With these two sets of model atmospheres we have computed the 
cooling sequences for the following two cases: 1) crystallization
and no phase separation and 2) crystallization and phase separation.
The results are shown in figure 2 and table 1. The left panel of 
figure 2 shows the cooling sequences for the non-DA model envelopes 
BH (solid lines) and MW (dotted lines). The sequences with phase 
separation (BH2 and MW2, respectively) correspond, obviously, to 
the cooling curves with larger cooling times for the same luminosity.
The right panel of figure 2 shows the same set of calculations for 
hydrogen-dominated white dwarf envelopes. We emphasize that these
figures highlight the {\sl relative} time delays due to chemical 
differentiation at crystallization which are obtained with either 
Wood (1995), Wood \& Winget (1989) or Hansen (1999) effective 
temperature-interior temperature relationships. 

\placetable{table1}

Finally in table 1 we show the cooling times (in Gyr) for the eight 
cooling sequences described above at the approximate position of the 
observed turn-off of the white dwarf luminosity function, $\log(L/
L_{\sun})\simeq -4.5$. From a superficial analysis of the data 
shown in table 1 it can be concluded that although the absolute 
value of the delay introduced by chemical segregation changes
appreciably due to the transparency of the envelope, the relative
contribution remains roughly constant and of the order of 10\% for
$\log (L/L_\odot)\approx -4.5$, in agreement with our previous 
results, which is not at all negligible even in an astrophysical 
context. This behavior is easily explained by the fact that the 
lifetime of white dwarfs in the cut-off of the white dwarf luminosity 
function is dominated by the crystallization phase and any change in 
the luminosity has the same influence on all the contributing sources 
of energy. It is worth mentioning that these results, similar to the
ones obtained in Segretain et al. (1994) and Hernanz et al. (1994), 
have been confirmed recently by Montgomery et al. (1999) who computed 
white dwarf cooling sequences, including the afore-described complete 
treatment of crystallization, with a complete stellar evolutionary 
code. Finally, it should be mentioned as well that if a flat initial profile 
is assumed for the C/O mixture, then the delay for a 0.606 $M_{\sun}$
white dwarf amounts to 1.8 Gyr when the DA envelope of Hansen (1999) is 
adopted (i.e., a $\sim$20\% effect), and 0.8 Gyr when the non--DA envelope 
is assumed.

\section{Conclusions}

We have derived the energy release due to the chemical 
redistribution associated to the process of crystallization of
C/O white dwarfs. We reassert that this contribution cannot
be by any means neglected, in contrast with the recent 
calculations of Hansen (1999). We have demonstrated that 
the reason of this discrepancy stems from an incorrect caculation 
of the aforementioned energy, due to a missing (molecular weigth) 
factor. This led Hansen (1999) to claim that the bulk of the 
correction in the cooling times introduced by the redistribution
of C/O upon crystallization was just the consequence of using an 
inaccurate envelope model. Although the importance of a correct 
envelope model to determine absolute ages is certainly crucial, 
meaningful comparisons of the different energy sources can only 
be done on the basis of relative contributions for the same kind 
of models (that is, the same envelopes). It is evident that any 
increase of the transparency of the atmosphere will globally reduce 
the time scales of white dwarf cooling, including the delay 
introduced by crystallization. When meaningful comparisons are 
done, we have shown that the contribution of chemical segregation 
to the duration of the crystallization phase is at least equivalent 
to the contribution of the latent heat itself and strongly depends 
on the initial chemical profile of the white dwarf. Therefore, the 
decision of neglecting or not this phenomenon only depends on the 
demanded degree of accuracy in the models. For instance, should we 
want to determine the age of the Galaxy or the ages of globular 
clusters with a precision of tenths of Gyr this effect should be 
included. 

Finally, it should be mentioned as well that the chemical profiles 
of Salaris et al. (1997) used here show a notoriously high abundance 
of oxygen in the central layers, which is caused by an enhanced 
$^{12}$C$(\alpha,\gamma)^{16}$O effective reaction rate, which is 
still the subject of a strong debate. This, of course, minimizes 
the effect of chemical diferentiation upon crystallization. 

\vskip 2cm

\noindent
{\em Acknowledgements}
This work has been supported by the DGICYT grants PB97--0983--C03--02 
and PB97--0983--C03--03 and by the CIRIT. We also want to thank Brad 
Hansen for sending us his $L-T_{\rm c}$ relationships wich allowed 
meaningful comparisons.

\newpage

\figcaption[fig1.ps]{The relationships between the core temperature
and the luminosity for the various model atmospheres adopted in this
work, see text for details
\label{fig1}}

\figcaption[fig2.ps]{Cooling curves (time is in Gyr) for the white 
dwarf models described in the text and in Table 1, the thinner 
horizontal line corresponds to $\log(L/L_{\sun})=-4.5$, which is
the approximate position of the observed cut-off of the white 
dwarf luminosity function.
\label{fig2}}

\newpage

\begin{deluxetable}{cccccc}
\tablecaption{Cooling times (in Gyr) at $\log(L/L_{\sun})=-4.5$ for the 
eight models studied in this paper. See text for details.\label{table1}}
\tablehead{
\colhead{Model} &
\colhead{DA} &
\colhead{non-DA} &
\colhead{Model} &
\colhead{DA} &
\colhead{non-DA} 
}
\tablewidth{0 pt}
\startdata
BH1 &7.92 &5.09 &MW1 &7.53 &8.83 \nl
BH2 &8.91 &5.70 &MW2 &8.41 &9.89 \nl
\enddata
\nl
\end{deluxetable}

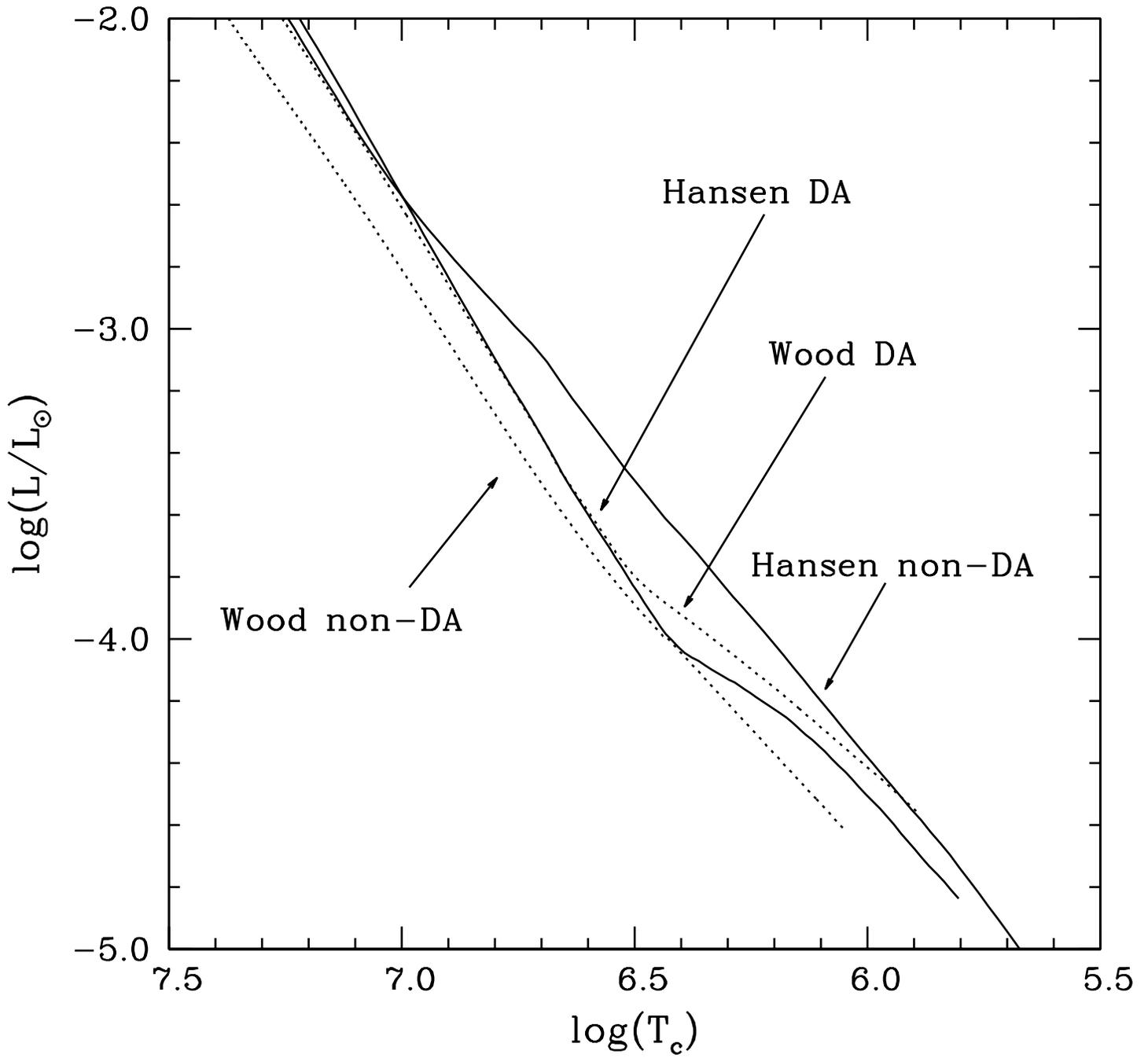
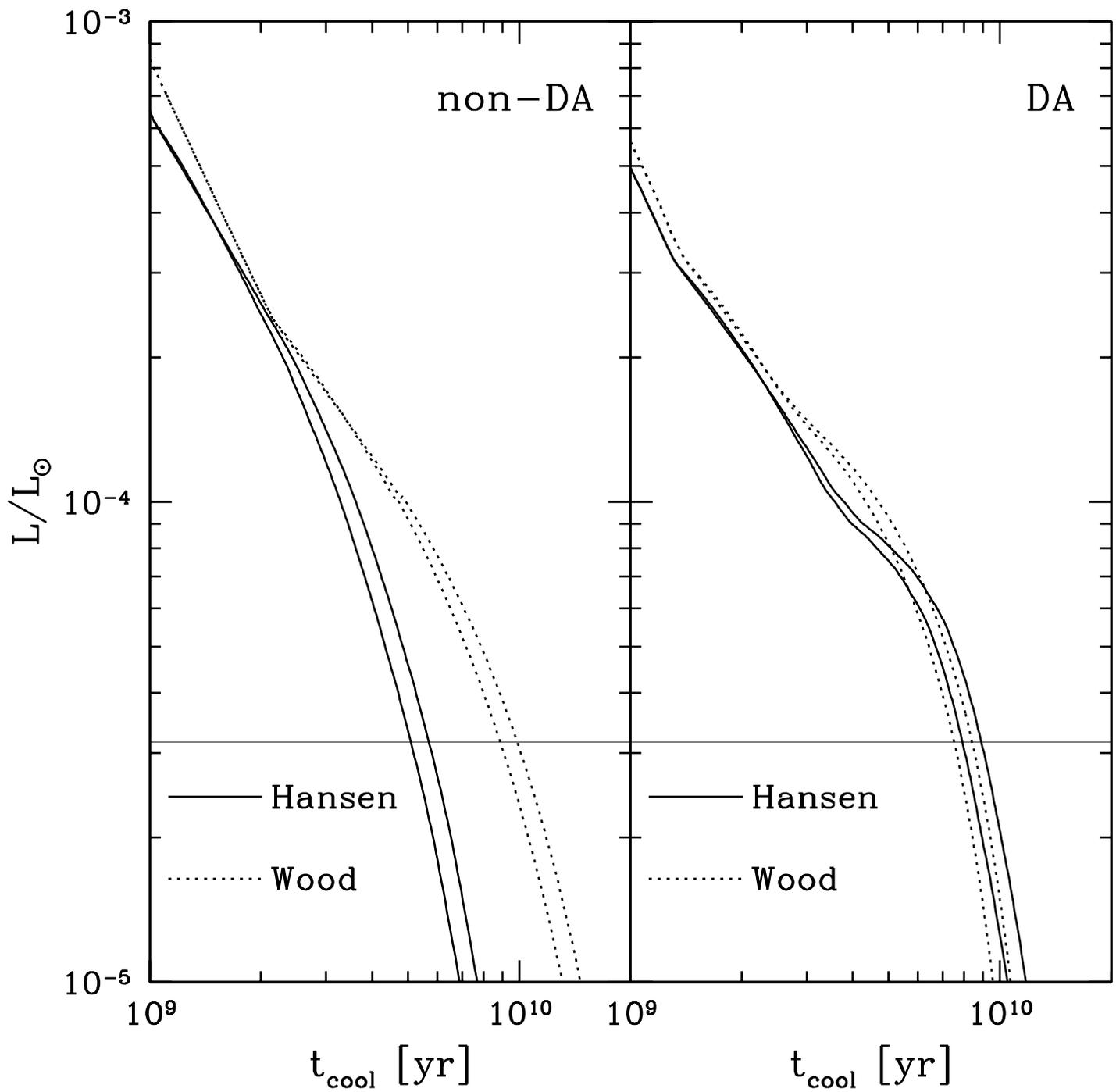
\end{document}